# Orientational ordering of confined hard rods: the effect of shape anisotropy on surface ordering and capillary nematization

R. Aliabadi[1,*], M. Moradi[1,†], S. Varga[2,‡]

[1] *Department of Physics, College of Science, Shiraz University, Shiraz 71454, Iran*
[2] *Institute of Physics and Mechatronics, University of Pannonia, PO Box 158, Veszprém, H-8201 Hungary*

We examine the ordering properties of rectangular hard rods with length $L$ and diameter $D$ at a single planar wall and between two parallel hard walls using the second-virial density functional theory. The theory is implemented in the three-state Zwanzig approximation, where only three mutually perpendicular directions are allowed for the orientations of hard rods. The effect of varying shape anisotropy is examined at $L/D = 10, 15$ and $20$. In contact with a single hard wall, the density profiles show planar ordering, damped oscillatory behavior, and wall induced surface ordering transition below the coexisting isotropic density of bulk isotropic-nematic (*I-N*) phase transition. Upon approaching the coexisting isotropic density, the thickness of nematic film diverges logarithmically, i.e. the nematic wetting is complete for any shape anisotropy. In the case of confinement between two parallel hard walls, it is found that the continuous surface ordering transition depends strongly on the distance between confining walls (*H*) for $H < L$, while it depends weakly on $H$ for $H > L$. The minimal density at which a surface ordering transition can be realized is located at around $H \sim 2D$ for all studied shape anisotropies due to the strong interference effect between the two hard walls. The first order *I-N* phase transition of the bulk system becomes surface ordered isotropic ($I_B$)-capillary nematic ($N_B$) phase transition in the slit pore. This first order $I_B$ - $N_B$ transition weakens with decreasing pore width and terminates in a critical point for all studied shape anisotropies.

Number of pages: 22 (including figure captions and figures)

Number of figures: 7

---

[*] r_aliabadi@shirazu.ac.ir
[†] moradi@susc.ac.ir
[‡] vargasz@almos.uni-pannon.hu

# I. INTRODUCTION

In the early 1940s, Onsager presented a theoretical verification that the anisotropic hard body interactions are sufficient to the formation of nematic phase when the density is sufficiently large [1]. His theoretical description was encouraged by the experimental observations of isotropic-nematic (*I-N*) phase separation in systems of $V_2O_5$ (ribbon-like vanadium pentoxide) [2] and rod-like tobacco mosaic virus (*TMV*) particles [3]. Later such transition was also discovered in other suspensions with different particle shapes, for instance, board-like goethite ($\alpha-FeOOH$) particles [4]. In the Onsager theory the Taylor expansion of the Helmholtz excess free energy functional is truncated at the second order and becomes exact for very long hard rods at low concentration. The theory predicts correctly that the bulk *I-N* transition to be of first order [5].

Due to existence of much richer phase behavior in confined liquid crystals, the surface effects are very interesting in practical applications. For example, the walls can give rise to different anchoring phenomena and result in various phase behaviors like a shift in the phase transition with respect to that of the bulk phase [6, 7]. In general, when a liquid is located in contact with another phase, like a solid surface, the substrate gives rise to substantial changes in the structure and the phase behavior of the liquid [8, 9]. This is because of the strong competition between particle-particle (intrinsic interactions) and particle-substrate (extrinsic interactions) interactions [10]. This is especially true for mesophase forming systems, where the solid surface may change the isotropic or nematic fluid structure from uniaxial to biaxial. It is predicted that the wall induced uniaxial-biaxial (*U-B*) phase transition is continuous for confined long hard rods [11, 12]. However, even first order *U-B* transition can be also found in the system of semiflexible polymers with hard core interactions [13].

Over the last decade much experimental [14, 15] and theoretical [11, 16, 17] works have supported the progress in the field of biaxial nematic phase (first prediction by Freiser [18]). The formation of a biaxial nematic film with the director parallel to the wall has been notified in colloidal hard-rod fluids near a single wall and confined between two hard walls using both theory and simulation [11, 12, 19]. The film wets the surface completely as the density approaches the density of the bulk *I-N* transition [11, 12, 20]. Complete wetting requires that the wall-isotropic (*WI*) surface tension to be the sum of the wall-nematic (*WN*) and *I-N* surface tensions. This will be discussed in the results section. It has been also shown that the first order capillary nematization arising from confining walls, where the *I-N* phase transition occurs at the lower chemical potential with respect to the bulk value [21], terminates in a capillary critical point with narrowing the width of the pore. The above results can be summarized as follows: 1) the wall induces a surface transition from uniaxial to biaxial symmetry, 2) the nematic film wets completely the wall-isotropic fluid interface, and 3) the critical wall separation at which capillary nematization terminates about twice the length of the rod [11]. Note that similar behavior has been also seen in the system of two-dimensional hard rods in some geometric confinements [22].

In this paper, we concentrate theoretically on the effects of shape anisotropy of rectangular hard rods with aspect ratios ($L/D$) 10, 15 and 20 on the *I-N* interfacial tension of the *I-N* interface, on the surface ordering and nematic wetting against a single hard wall, and on the capillary nematization and surface ordering between two parallel hard walls. In section 2 we briefly present the Onsager theory within Zwanzig approximation [23] where only three mutually perpendicular directions are allowed. Even though this model is simple, it captures the basic physics of the systems and also makes the numerical analysis substantially easier [11]. In section 3 we present our numerical results and discussion. The conclusion is given in section 4 and finally we show a new method for the location of *U-B* transition in the appendix.

## II. THEORY

We study the ordering properties of rectangular hard rods with length *L* and diameter *D* at the presence of confining hard walls using the second virial theory (or Onsager theory) and the so-called three-state Zwanzig approximations for the particles' orientations. In the Zwanzig approximation, particles' principle axes are allowed to point along one of the Cartesian axes (*x, y* and *z*) as shown in Fig. 1, i.e. the system corresponds to a ternary mixture of hard bodies without rotational freedom. In this three-state model the grand potential of the system on the level of Onsager theory can be written as

$$\beta\Omega = \sum_{i=x,y,z}\int d(1)\rho_i(1)\ln\rho_i(1) - \rho_i(1) - \frac{1}{2}\sum_{i,j=x,y,z}\int d(1)\rho_i(1)\int d(2)\rho_j(2)f_M^{ij}(1,2) \\ + \sum_{i=x,y,z}\int d(1)\rho_i(1)[\beta V_{ext}^i(1) - \beta\mu],$$
(1)

where $\beta = 1/(k_B T)$, $\rho_i$ is the local density of component *i* (long principle axes of particles are parallel with *i* axis (*i*=x,y,z)), $f_M^{ij}$ is the Mayer function between components *i* and *j*, and $V_{ext}^i$ is the external potential for component *i* and $\mu$ is the chemical potential. The integrations are performed in positions, i.e. $(i) = d\vec{r}_i$, where $\vec{r}_i = (x_i, y_i, z_i)$ is the position vector ($i=1,2$). Since the normal of the confining wall is parallel to the $z$ axis ($V_{ext}^i$ is $z$ dependent) and we are only interested in orientational ordering transitions, the local density components ($\rho_i$ ($i=x,y,z$)) depend only on the perpendicular distance ($z$) from the wall. Using the fact that $f_M^{ij}$ is equal to -1 for overlapping particles and zero otherwise, we can integrate out the in-plane variables ($x$ and $y$) in Eq. (1), i.e. the grand potential simplifies to

$$\beta \Omega / A = \sum_{i=x,y,z} \int dz \rho_i(z) \ln \rho_i(z) - \rho_i(z) + \frac{1}{2} \sum_{i,j=x,y,z} \int dz_1 \rho_i(z_1) \int dz_2 \rho_j(z_2) A_{exc}^{ij}(z_{12})$$
$$+ \sum_{i=x,y,z} \int dz_1 \rho_i(z_1) [\beta V_{ext}^i(z_1) - \beta \mu], \qquad (2)$$

where $A$ is the surface of the confining walls, $A_{exc}^{ij}(z) = -\int dx \int dy f_M^{ij}$ is the excluded area between two hard particles with orientations $i$ and $j$ ($i,j=x,y,z$)) and $z_{12} = z_1 - z_2$ is the vertical distance between two rods, respectively. For rectangular hard rods with length $L$ and cross section lengths $D$, the excluded areas can be determined analytically which are given by

$$A_{exc}^{xx}(z) = A_{exc}^{yy}(z) = 4DL \quad for -D < z < D,$$

$$A_{exc}^{zz}(z) = 4D^2 \quad for -L < z < L,$$

$$A_{exc}^{xy}(z) = A_{exc}^{yx}(z) = (L+D)^2 \quad for -D < z < D,$$

and

$$A_{exc}^{xz}(z) = A_{exc}^{zx}(z) = A_{exc}^{yz}(z) = A_{exc}^{zy}(z) = 2D(L+D) \quad for -\frac{L+D}{2} < z < \frac{L+D}{2}. \qquad (3)$$

Note that the above excluded areas are symmetric, i.e. $A_{exc}^{ij} = A_{exc}^{ji}$, and zero if the $z$ distance between the two particles is out of the indicated intervals. The external potential of the system confined between two parallel walls is purely hard and defined for parallel and perpendicular particles' orientations to the walls as follows:

$$\beta V_{ext}^i(z) = \begin{cases} \infty, & z < D/2 \text{ and } z > H - D/2, \quad (i=x, y) \\ 0, & D/2 < z < H - D/2 \end{cases}$$

$$\beta V_{ext}^{z}(z) = \begin{cases} \infty, & z < L/2 \text{ and } z > H - L/2, \\ 0, & L/2 < z < H - L/2 \end{cases} \qquad (4)$$

where $H$ is the distance between the two parallel walls. In the case of single wall, since the wall is located at $z=0$ the external potential is infinite only for distances below $D/2$ and $L/2$ in the above expressions, respectively. The equilibrium local densities are obtained from the functional minimization of the grand potential with respect to the local densities, i.e. $\frac{\delta \beta \Omega / A}{\delta \rho_k(z)} = 0$. The resulting three integral equations can be written concisely as

$$\rho_i(z) = \exp\left\{ - \sum_{j=x,y,z} \int dz_1 \rho_j(z_1) A_{exc}^{ij}(z - z_1) \right\} \exp\left\{ -\beta V_{ext}^i(z) \right\} \exp\{\beta \mu\}, \quad (i = x, y, z) \qquad (5)$$

It can be seen that Eq. (5) corresponds to three self-consistent and coupled equations for the local densities $\rho_x$, $\rho_y$ and $\rho_z$ at a given chemical potential ($\mu$). We have solved Eq. (5) by the standard iteration method and used the trapezoidal quadrature for the calculations of the integrals. Although $x$ and $y$ orientations are equivalent, we have chosen such trial distribution functions for the local densities, which may result in such solutions that $\rho_x \geq \rho_y$. In the bulk phase, where the external potential term is missing, the above coupled equations become much simpler, because the local densities lose their positional dependence. From Eq. (5) one can derive that

$$\rho_i = \exp\left[ - \sum_{j=x,y,z} \rho_j V_{exc}^{ij} \right] \exp[\beta \mu], \quad (j = x, y, z) \qquad (6)$$

where $V_{exc}^{ij} = \int dz \, A_{exc}^{ij}(z)$ is the excluded volume between two rods. In this case only parallel and perpendicular orientations can be distinguished, i.e. $V_{exc}^{\parallel} = V_{exc}^{ii} = 8LD^2$ ($i = x, y, z$) and $V_{exc}^{\perp} = V_{exc}^{ij} = 2D(L+D)^2$ ($i \neq j$). In bulk one of the solutions of Eq. (6) is isotropic, i.e. $\rho_x = \rho_y = \rho_z$, while the other solution is nematic, where $\rho_x > \rho_y = \rho_z$. In the confined situation three different solutions can be obtained, one is isotropic, while the other two are orientationally ordered phases. To characterize the type and the degree of ordering we use three different orientational order parameters. One of them is the orientation average of the second Legendre polynomial if the director is chosen to be along $x$ axis i.e. $S_x = \langle P_2 \rangle_{\vec{n}=(1,0,0)}$, while the others order parameters ($\Delta_{xy}$ and $\Delta_{yz}$) measure the degree of biaxiality between two in-plane orientations (*x* and *y*) and between one in-plane and one out-of-plane orientations (*y* and *z*). In our approach the formulas for these order parameters are

$$S_x = \frac{\rho_x - \rho_y/2 - \rho_z/2}{\rho}, \tag{7}$$

$$\Delta_{xy} = \frac{\rho_x - \rho_y}{\rho}, \tag{8}$$

and

$$\Delta_{yz} = \frac{\rho_y - \rho_z}{\rho}, \tag{9}$$

where $\rho = \rho_x + \rho_y + \rho_z$ is the total local density. In the bulk isotropic phase all these order parameters are zero, while $0 < S_x < 1$ and one of the biaxial order parameters is zero (either $\Delta_{xy} = 0$ or $\Delta_{yz} = 0$) in the bulk nematic phase. In the case of confined systems the local densities are not constant and the confining walls induce planar ordering at the vicinity of the walls even for arbitrary small chemical potentials. Therefore the isotropic phase of confined systems has the property that $S_x \geq 0$ and only one biaxial order parameter is non-zero ($\Delta_{yz} > 0$ and $\Delta_{xy} = 0$) which corresponds to the case when $\rho_x = \rho_y > \rho_z$. The other phase is also isotropic to some extent, but surface ordering takes place at the vicinity of the walls, i.e. the phase is biaxial at the walls. In this case the local densities are non-equal at the vicinity of walls ($\rho_x > \rho_y > \rho_z$), but $\rho_x = \rho_y \geq \rho_z$ happens far from the walls. This corresponds to $S_x > 0$ and $\Delta_{xy}, \Delta_{yz} > 0$ close to the walls, while $\Delta_{xy} = 0$ far from the walls. The nematic phase is biaxial at the walls, i.e. $S_x > 0$ and $\Delta_{xy}, \Delta_{yz} > 0$, but it becomes differently uniaxial far from the walls due to $\rho_x > \rho_y = \rho_z$ ($\Delta_{xy} > 0$ and $\Delta_{yz} = 0$) if the pore is very wide. We present the local density and order parameter curves of different structures in the next section. Since both first and second order phase transitions may occur in confined systems, we undertake a bifurcation analysis between uniaxial and biaxial phases and determine the coexisting densities of different phases. In our case the bifurcation analysis, which is presented in the Appendix, gives the phase boundary of the surface ordering phase transition since this transition is of second order. The properties of the first order phase transitions both in bulk and confined cases, which takes place between isotropic and nematic phases, are determined by searching the cross point between two different solutions of Eq. (5) in $\beta\Omega/A - \beta\mu$ plane.

The calculated equilibrium density profiles can be also used to determine the interfacial surface tensions ($\gamma_{WI}, \gamma_{WN}$ and $\gamma_{IN}$). These profiles are inserted into Eq. (2) to get the equilibrium value of the grand potential ($\Omega_0$) of the system. All surface tensions have been calculated using the general definition of the surface tension as the surface excess grand potential per unit area

$$\gamma = \frac{\Omega_0 + pV}{A} \tag{10}$$

where $p$ is the pressure and $V$ is the volume of the system. In the figures dimensionless number densities ($\rho^* = \rho D^3$), pore width ($h^* = H/D$) and vertical position $z^* = z/D$ are utilized.

### III. RESULTS AND DISCUSSION

Prior to study of the structure and phase behavior of the confined hard rod fluid we have calculated the homogeneous bulk *I-N* phase boundary, as shown in Fig. 2, by searching the cross point between two different solutions of Eq. (6) in $\beta\Omega/A - \beta\mu$ plane. Fig. 2 shows that the *I-N* phase transition is weakly first order and the coexisting densities decrease with increasing shape anisotropy. In the inset of Fig. 2 one can see that the orientational order parameter of the coexisting nematic phase increases with increasing $L/D$. Apart from the shift in the coexisting densities, these results are in good agreement with the theoretical and simulation results obtained for freely rotating hard rods. This means that the three-state Zwanzig approximation represents correctly the orientational freedom of the particles. Using Fig. 2 one can see the reference values for our confined results in the interval $10 < L/D < 20$.

In Fig. 3 we exhibit density profiles and the corresponding order parameters of the particles with $L/D = 10$ in contact with a single hard wall for three different chemical potentials where the hard planar wall is located at $z^* = 0$. The presented density profiles show sharp and high first peak with a large contact value of $\rho_x^*$ for all chemical potentials, i.e. there is a strong adsorption at the wall. In addition to this the density profiles exhibit damped oscillatory behavior near the planar wall, while they converge to the bulk density values far from the wall. Near the surface $\rho_z^*$ is always smaller than $\rho_b^*/3$, while $\rho_x^*$ is always higher at any distance for all chemical potentials shown. This shows that the ordering is planar at the wall. The phase is isotropic with planar order at $\beta\mu = -2.8$ as it can be seen from Fig. 3 (a) and (b) where $\rho_x^*$ is equal to $\rho_y^*$, i.e. surface induced biaxiality is not present ($\Delta_{xy} = 0$). Enough far from the wall all densities approach to the bulk density ($\rho_b^*/3 = 4.584 \times 10^{-3}$). As the chemical potential is increased the biaxial order starts to exist as the number of particles parallel with $x$ axis rises more than that of particles parallel with $y$ axis. In Fig. 3 (c) and (d), $\rho_x^* \neq \rho_y^*$ and $\Delta_{xy}$ as well as $\Delta_{yz}$ are nonzero spontaneously near the wall so the phase is biaxial. These figures show surface ordering and formation of a nematic film. In Fig. 3 (e), where the chemical potential is chosen to be close to $\beta\mu_{IN}^b \approx -2.1327$, an substantial increase in the value of $\rho_x^*$ can be seen with respect to $\rho_y^*$ and $\rho_z^*$. This indicates that the thickness of the nematic layer can be grown by increasing the chemical potential. This fact is also clear from the order parameters presented in Fig. 3 (f). There is a very weak negative oscillation in $\Delta_{yz}$ where $\rho_y^* < \rho_z^*$ in the Fig. 3 (f) right after $z^* = 5$ that becomes slightly positive before decaying quickly to zero ($\rho_y^* = \rho_z^*$). Furthermore $\Delta_{yz}$ becomes smaller for larger chemical potentials, which means that the values of $\rho_z^*$ and $\rho_y^*$ are getting closer to each other. At very large distances all densities become identical, i.e. the phase is isotropic far from the wall. The wall induced *U-B* transition occurs at $\beta\mu = -2.7550$ which is below the chemical potential of bulk *I-N* transition ($\beta\mu_{IN}^b$). Upon getting close to $\beta\mu_{IN}^b$, the thickness of nematic film diverges logarithmically. It is important to note that the nematic wetting is complete for any shape anisotropy as it is proved by the results for surface tensions. In summary Fig. 3 shows that how the second order biaxial surface ordering transition evolves with increasing chemical potential.

At some aspect ratios, the surface tensions have been calculated using Eq. (10) by choosing very large distance between the wall and the bulk fluid and between coexisting isotropic and nematic phases ($\sim 1000D$). In the case of wall-isotropic surface tension calculations, we have performed extrapolation as follow $\gamma_{IW}(\mu_{IN}^b) = \lim_{\mu \to \mu_{IN}^b} \gamma_{IW}(\mu)$ because the thickness of the nematic film diverges at $\beta\mu = \beta\mu_{IN}^b$. The result surface tensions are fitted by R-square larger than 0.98 to a power law, $\gamma^*(x) = ax^b$, where $x = L/D$. Fig. 4 presents the dimensionless wall-isotropic ($\gamma^*_{WI} = \beta\gamma_{WI}D^2$), wall-nematic ($\gamma^*_{WN} = \beta\gamma_{WN}D^2$) and I-N ($\gamma^*_{IN} = \beta\gamma_{IN}D^2$) surface tensions as a function of $L/D$, which are given by the following fitting parameters:

$\gamma^*_{WI}$ : $a = 1.2033441326$    $b = -1.7801894801$

$\gamma^*_{WN}$ : $a = 1.4847697494$    $b = -1.9088064481$

$\gamma^*_{IN}$ : $a = 0.0081512312$    $b = -0.6966685124$

Note that our fitting is not valid for infinitely long rods. Our results prove that the nematic wetting of the wall-isotropic (WI) fluid interface is complete by a nematic film, where the nematic director of the film is oriented parallel to the wall (x axis in our calculation). Applying above parameters, one can find that $\gamma_{IN}$ is equal to $\gamma_{WI} - \gamma_{WN}$. Therefore the contact angle must be zero which is apparent from the Young's equation, $\gamma_{IN}\cos\theta = \gamma_{WI} - \gamma_{WN}$. One can also show that the excess adsorption diverges logarithmically at $\beta\mu = \beta\mu^b_{IN}$, which is a feature of complete wetting of classical fluids interacting with short range forces [24]. Similar behavior is reported in the MC simulation study of hard spherocylinders [19]. It is also consistent with the profiles displayed in Fig. 3, which shows an increase in the thickness of the film with increasing chemical potential. As discussed before, the wetting transition is second order.

The density profiles and order parameters of the confined fluid by two planar hard walls are based on the numerical solution of the Euler-Lagrange equation (Eq. (5) combined with Eq. (4)). The results are presented in Fig. 5 for $L/D = 10$ at three different chemical potentials. In connection with single hard wall density profiles (Fig. 3), here it is also obvious that uniaxial phase changes to biaxial and most of the particles align parallel with x axis by increasing the chemical potential. Therefore, a biaxial nematic layer forms in the whole range of the pore. Here both walls act on the same way and support the formation of the nematic film with oscillatory density profiles. At some separations both walls strengthen the biaxial order (they are in-phase) but they can weaken it at other separations (they are out of phase). Therefore the appearance of the biaxial order at the vicinity of the walls is the result of the interference of the two walls. As it can be seen from Fig. 5 (a) and (b), the phase is isotropic at $\beta\mu = -3.48$ where $\rho^*_x$ and $\rho^*_y$ are equal even close to the walls and hence $\Delta_{xy} = 0$ while $\Delta_{yz} \neq 0$ and $S_x \neq 0$ except in the middle of the pore. The phase is biaxial isotropic (or more precisely weakly nematic) for $\beta\mu = -2.51$ due to $\rho_x^*$ and $\rho_y^*$ are different near the surfaces but equal in the middle of the pore (Fig. 5 (c)). In this phase, apart from the central region, $\Delta_{xy} \neq 0$ but $\Delta_{yz}$ and $S_x$ are non-zero everywhere (Fig. 5 (d)). Moreover, $S_x$ shows a very weak nematic ordering in the middle of the pore. The phase in Fig. 5 (e) is nematic biaxial ($\beta\mu = -2.00$) since $\rho_x^* > \rho_y^*$ and most of the particles are aligned in the direction of $x$ axis. The nematic film persists throughout the central region. In this phase, the order parameters $\Delta_{xy}$ and $S_x$ have large values whereas $\Delta_{yz}$ is very small due to the small difference between $\rho_y^*$ and $\rho_z^*$ (Fig. 5 (f)).

Based on the numerical solution of Eq. (5) we have found both biaxial isotropic and biaxial nematic solutions at some values of the chemical potential, which indicates that first order phase transitions occur in the slit-like pore. The resulting first order phase transitions between biaxial isotropic and nematic phases ($I_B$ - $N_B$) and the surface ordering transitions (U-B) are presented in a common phase diagram for different particle's shapes ($L/D = 10, 15$ and 20) in Fig. 6. In this figure the average density in the slit-pore, $\rho_{av}^* = \frac{1}{H}\int_0^H \rho^*(z)dz$, is shown as function of the reduced pore width ($h^* = H/D$). U-B transition is always second order like the one-wall case for any shape anisotropy and oscillates with varying pore width at short pore widths. This means that this transition strongly depends on the wall-to-wall distance if $h^* < L/D$. For more elongated particles, the heights of the waves get smaller and the transition occurs at smaller densities. When $h^* > L/D$, the U-B line starts to get smoothly varying and approaches the single wall limit of the surface ordering transition. This is due to the fact the walls are too far from each other and the interference weakens between them. The first order I-N transition of the bulk system becomes surface ordered $I_B$ - $N_B$ phase transition in the slit pore. This first order $I_B$ - $N_B$ transition weakens with decreasing pore width and terminates in a critical (capillary) point $h_c^*$ for all studied shape anisotropies similar to the studies for infinitely long hard rods [11, 12] and finite rods [19]. As a result the two coexisting biaxial phases get fully indistinguishable as $h^* \to h_c^{*+}$. $N_B$ and $I_B$ phases are less and less biaxial far from the walls if $H$ goes to infinity.

$I_B$-$N_B$ coexistence densities in the pores are smaller in comparison with their bulk *I-N* densities (see Fig. 2) because the planar walls promote the formation of nematic ordering. These coexisting densities of the confined system are approaching the bulk *I-N* coexisting densities with increasing pore width. The critical point for particles with $L/D=10$ and 15 are located at $h_c^* = 32.1605$ and 43.1630 respectively. Similar result has been reported for a fluid of hard spherocylinders with a length-to-diameter ratio of 15 by a simulation study [19]. The critical point for particles with $L/D=20$ is close to $h_c^* \approx 54$ (we have not managed to determine it more accurately). Using $L$ as a unit of distance and $\rho_{av} L^2 D$ as a dimensionless density, then the transition densities of $L/D=10,15$ and 20 cases become in the same order of magnitude and the phase diagrams can be shown together in $\rho_{av} L^2 D - H/L$ plane (see Fig. 6 (d)). The main advantage of this representation is that it is possible to make a connection with the well-known Onsager-limit ($L/D \to \infty$). One can see that the *U-B* surface ordering transition becomes less oscillatory and it is shifted into the direction of the limiting value of $\rho_{av} L^2 D \approx 1.031$, which is obtained for $L/D \to \infty$ case [11,12]. It can be also seen that the critical pore-width of $I_B$-$N_B$ transition is getting closer to the limiting value of $H_c/L=2.08$ with increasing aspect ratio. Fig. 7 shows the relationship between the calculated $I_B$-$N_B$ critical points and $D/L$. To see the correction of the finite aspect ratio we make the following fit for the critical pore-width: $H_c/L = 2.08(1 + a x^b)$, where $x = D/L$. This formula reproduces the $L/D \to \infty$ case exactly, where $H_c/L = 2.08$. One can see that our fitting formula with $a = 4.24875$ and $b = 0.89038$ values reproduces quite accurately the numerical results. The reason why even more than 50% difference emerges between $H_c/L$ of $L/D=10$ and $L/D \to \infty$ cases is that $A_{exc}^{ii}$ terms are not negligible and getting more dominant at lower aspect ratios.

### IV. SUMMARY AND CONCLUSION

In this work we have studied the influence of shape anisotropy of rectangular hard rods on the structural properties and phase behavior of isotropic and nematic phases in the presence of confinement using the well-known Onsager's second virial theory with three-state restriction for the orientations of the hard rods (Zwanzig approximation). Fluids of hard rods with shape anisotropies of $L/D=10,15$ and 20 have been considered in contact with a single planar hard wall and in a confinement where the rods are constrained to stay between two parallel hard walls. In the first part of the study we have calculated bulk coexisting *I-N* densities as a function of $L/D$. It is found that the Zwanzig approximation, which is widely used for both bulk [25] and confined systems [26], does not change qualitatively the results for the *I-N* coexisting densities and order parameters, i.e. it shows decreasing transition densities and increasing order parameters with increasing $L/D$ in agreement with the theoretical results obtained for freely rotating hard rods. In the case of single hard wall, we have found that the uniaxial phase changes to the biaxial one near the wall as the chemical potential is increased. Wall induced planar surface ordering and the formation of biaxial nematic film parallel to the wall have been observed for all shape anisotropies. The planar ordering is thermodynamically more favorable than the homeotropic ordering, where the nematic director is perpendicular to the wall. The thickness of the nematic film diverges logarithmically, as it is expected for systems interacting with short-range forces. The calculated three surface tensions are fitted greatly by asymptotic functions. They show a zero contact angle which means that the nematic layer wets the wall-isotropic interface completely for any shape anisotropy.

In the second part of this study we have considered the fluid of hard rods confined between two parallel hard walls. It is found that the increasing chemical potential stabilizes the biaxial nematic film that is persisting throughout the pore. There is a continuous surface ordering transition from uniaxial to biaxial symmetry with smoothly varying transition density which is below the bulk *I-N* coexistence densities for $h^* > L/D$, while the surface ordering transition density oscillates and depends strongly on the wall-to-wall separation for $h^* < L/D$. The weak and the strong $h^*$ dependence of the surface ordering transition density is due to the weak and strong interference of the two wall induced orderings. We have found first order capillary nematization transition which terminates at a critical pore width $h_c^*$ for any shape anisotropy. Our study reveals the difference between the ordering properties of hard rods in narrow and wide pores. Only surface ordering transition takes place in narrow pores because the system is reminiscent of quasi-two-dimensional systems of hard rods, where the ordering transition is continuous [27]. In wider pores both surface ordering transition and capillary nematization are present due to the strong adsorption at the walls and the competition between packing and orientational entropies. These transitions occurs even at infinite pore width ($h^* \to \infty$), where the walls does not effect the *I-N* properties, i.e. the coexisting densities of the $I_B$ and $N_B$ phases become identical to those of bulk *I-N* transition. In comparison with previous theoretical results [11, 12], our study show that Onsager theory is capable to result in richer local structures for the confined isotropic and nematic phases and to describe the oscillatory behavior of the surface ordering transition if all terms of the excluded volume interactions are incorporated into the calculations.

As an extension of our present work we plan to study the ordering properties of board-shaped hard rods, where the cross section lengths are different ($D_1 < D_2 < L$). This allows us to study the phase behavior of confined goethite nanorods which is proven to be the first colloidal system showing biaxial nematic ordering [28-31]. Along this line the effects of shape polydispersity and the external magnetic field on the stability of biaxial nematic phase have been already the subject of recent theoretical and experimental studies [32,33].

**APPENDIX. BIFURCATION ANALYSIS OF SURFACE ORDERING TRANSITION**

The simplest method of finding the second order transition point between disordered and ordered phases at the surfaces is to decrease the chemical potential with fine steps. At high chemical potentials we obtain ordered solutions with $\Delta_{xy} \neq 0$ and $\Delta_{yz} \neq 0$ close to the walls, while $\Delta_{xy}$ becomes zero at the transition point. Instead of decreasing the chemical potential and looking at the value of $\Delta_{xy}$ when it becomes zero, we present here an elegant method for the location of surface ordering transition.

In the isotropic phase of confined hard particles the in-plane local densities are identical, while the out-of-plane one is different ($\rho_x(z) = \rho_y(z) \neq \rho_z(z)$). Assuming that the biaxial order evolves continuously from the isotropic one, we apply the following ansatz for the biaxial order

$$\hat{\rho}_x(z) = \rho_x(z)(1 + \varepsilon(z)),$$
$$\hat{\rho}_y(z) = \rho_y(z)(1 - \varepsilon(z)),$$
$$\hat{\rho}_z(z) = \rho_z(z), \qquad (A1)$$

where $\varepsilon(z)$ is an unknown biaxial perturbation. Note that $\rho_x(z)$, $\rho_y(z) = \rho_x(z)$ and $\rho_z(z)$ are the isotropic solution of Eq. (5) . Since Eq. (A1) also satisfy Eq. (5) we get that the biaxial perturbation obeys the following equation

$$1+\varepsilon(z) = \exp\left\{-\int_{\max(D/2,z-D)}^{\min(H-D/2,z+D)} dz'\varepsilon(z')\rho_x(z')A_{exc}^{xx} + \int_{\max(D/2,z-D)}^{\min(H-D/2,z+D)} dz'\varepsilon(z')\rho_x(z')A_{exc}^{xy}\right\}, \quad (A2)$$

where $A_{exc}^{xx} = 4LD$ and $A_{exc}^{xy} = (L+D)^2$. After linearization of the exponential function ($\exp(x) \approx 1+x$) and the rearrangement of Eq. (A2) we get that

$$\int_{\max(D/2,z-D)}^{\min(H-D/2,z+D)} dz'\varepsilon(z')\left\{\delta(z-z') + \left[A_{exc}^{xx} - A_{exc}^{xy}\right]\rho_x(z')\right\} = 0, \quad (A3)$$

where $\delta(z)$ is the Dirac delta function. Formally Eq. (A3) can be satisfied with $\varepsilon(z) \neq 0$ if the determinant of argument vanishes, i.e.

$$\det\left|\delta(z-z') + \left[A_{exc}^{xx} - A_{exc}^{xy}\right]\rho_x(z')\right| = 0. \quad (A4)$$

In practice, we have produced a matrix using the discretized values of the local density of the isotropic solution of Eq. (5) and we have searched for that chemical potential at which the local density satisfies both Eqs. (A4) and (6). Results coming from Eq. (A4) and the method of decreasing chemical potential agree perfectly. The main advantage of our method is that we do not have to determine the biaxial perturbation ($\varepsilon(z)$).

## Figures

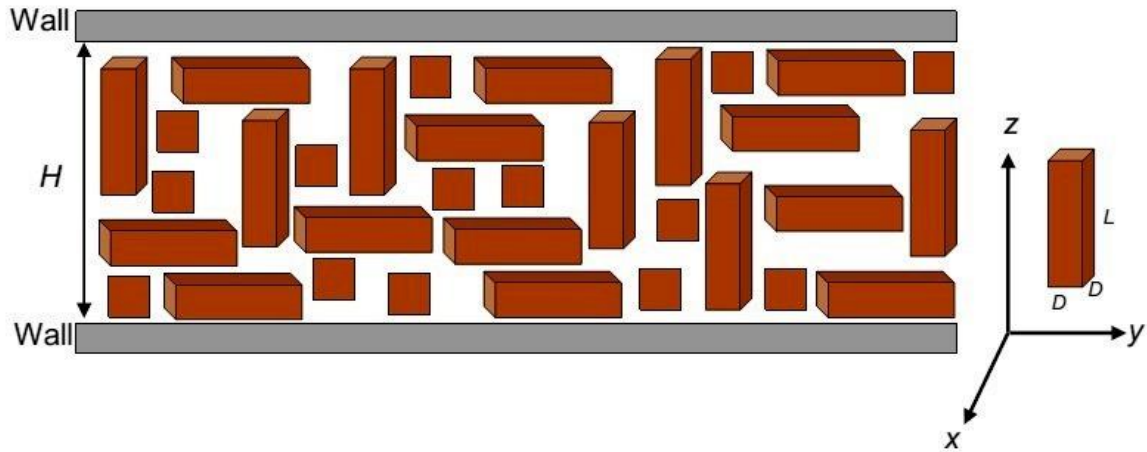

**FIG. 1.** (Color online) Schematic representation of the rectangular hard rods confined in a slit pore. Particles are allowed to orient along $x$, $y$ and $z$ axes (Zwanzig approximation). $H$ is the wall-to-wall separation. $L$ and $D$ are the length and the diameter of hard rods, respectively.

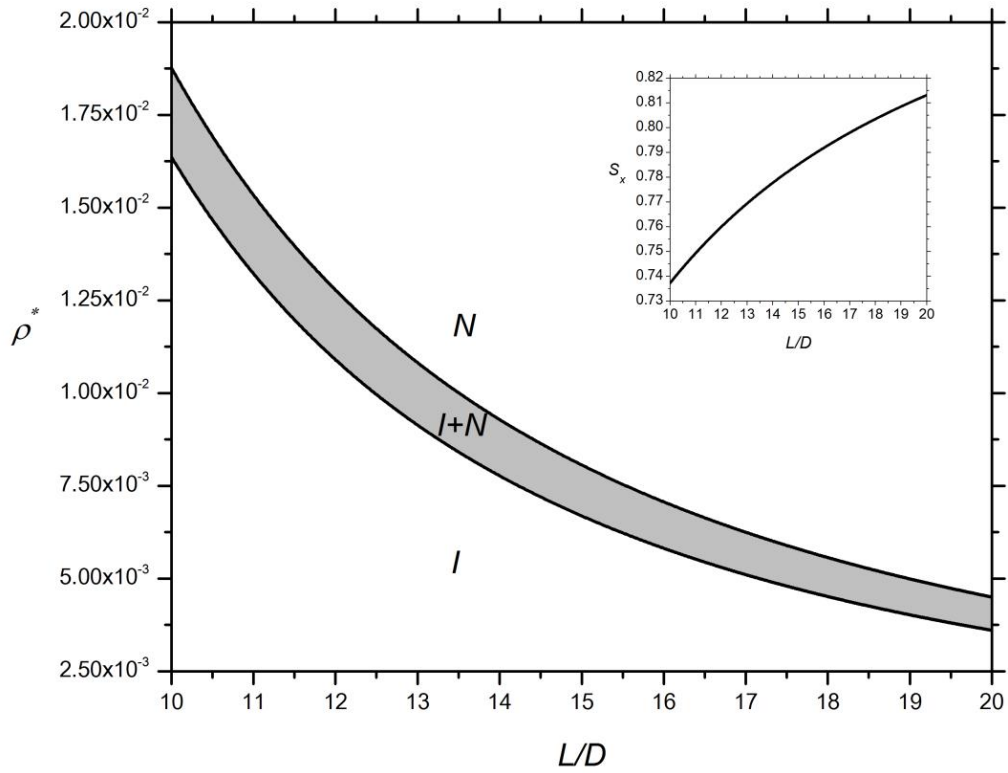

**FIG. 2.** Isotropic-Nematic phase boundary of hard rods in bulk. Coexisting isotropic (*I*) and nematic (*N*) densities are shown as a function of aspect ratio ($L/D$). Inset depicts the nematic order parameter ($S_x$) of the coexisting nematic phase. $\rho^*$ is the reduced number density ($\rho^* = \frac{N}{V}D^3$).

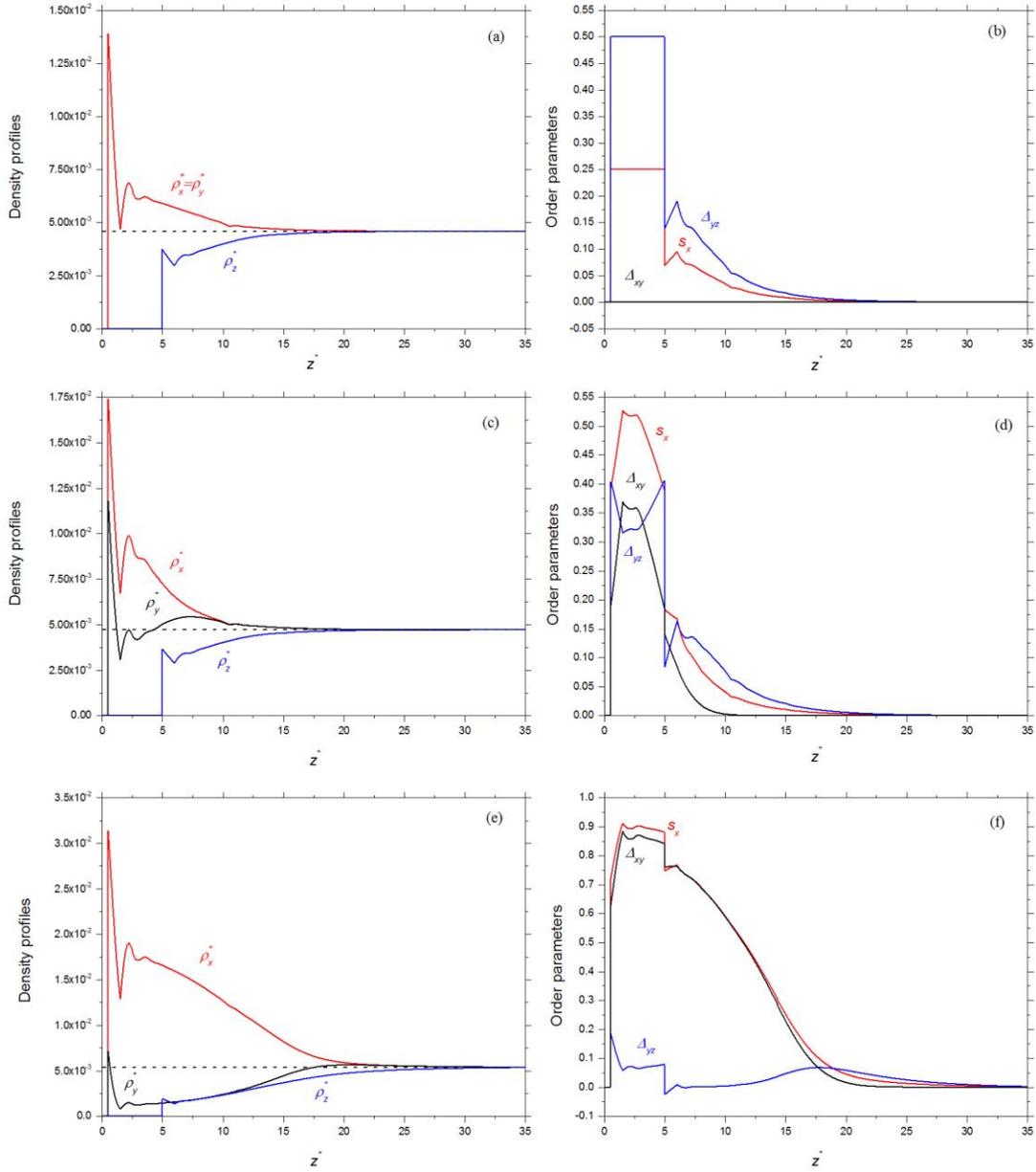

**FIG. 3.** (Color online) Density profiles and order parameters of the hard rod fluid in contact with a single hard wall which is located at $z^* = 0$. The aspect ratio ($L/D$) is chosen to be 10. Figs. (a) and (b) are obtained at $\beta\mu = -2.8$ where the phase is isotropic and the bulk component densities (dotted line) equal to $4.584 \times 10^{-3}$. Figs. (c) and (d) correspond to the case $\beta\mu = -2.7$. This phase is biaxial isotropic and the corresponding bulk component densities (dotted line) are $4.712 \times 10^{-3}$. In Figs. (e) and (f) the reduced chemical potential is -2.2, where the bulk component densities are $5.368 \times 10^{-3}$.

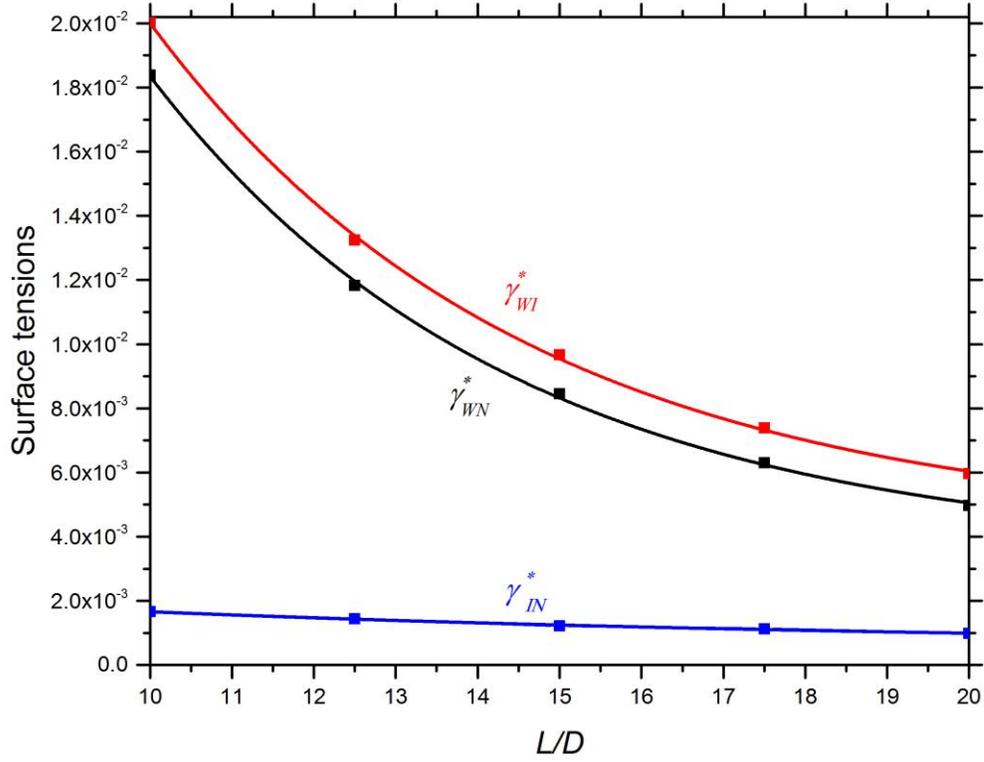

**FIG. 4.** (Color online) Fitted surface tensions $\gamma_{WI}$, $\gamma_{WN}$ and $\gamma_{IN}$ of the hard wall-isotropic, hard wall-nematic and *I-N* fluid interfaces respectively. The calculated surface tensions (solid squares) are fitted by power law, $\gamma(x) = ax^b$ (solid lines) where $x = L/D$. The surface tensions are dimensionless quantities, i.e. $\gamma_{WI}^* = \beta\gamma_{WI}D^2$, $\gamma_{WN}^* = \beta\gamma_{WN}D^2$ and $\gamma_{IN}^* = \beta\gamma_{IN}D^2$.

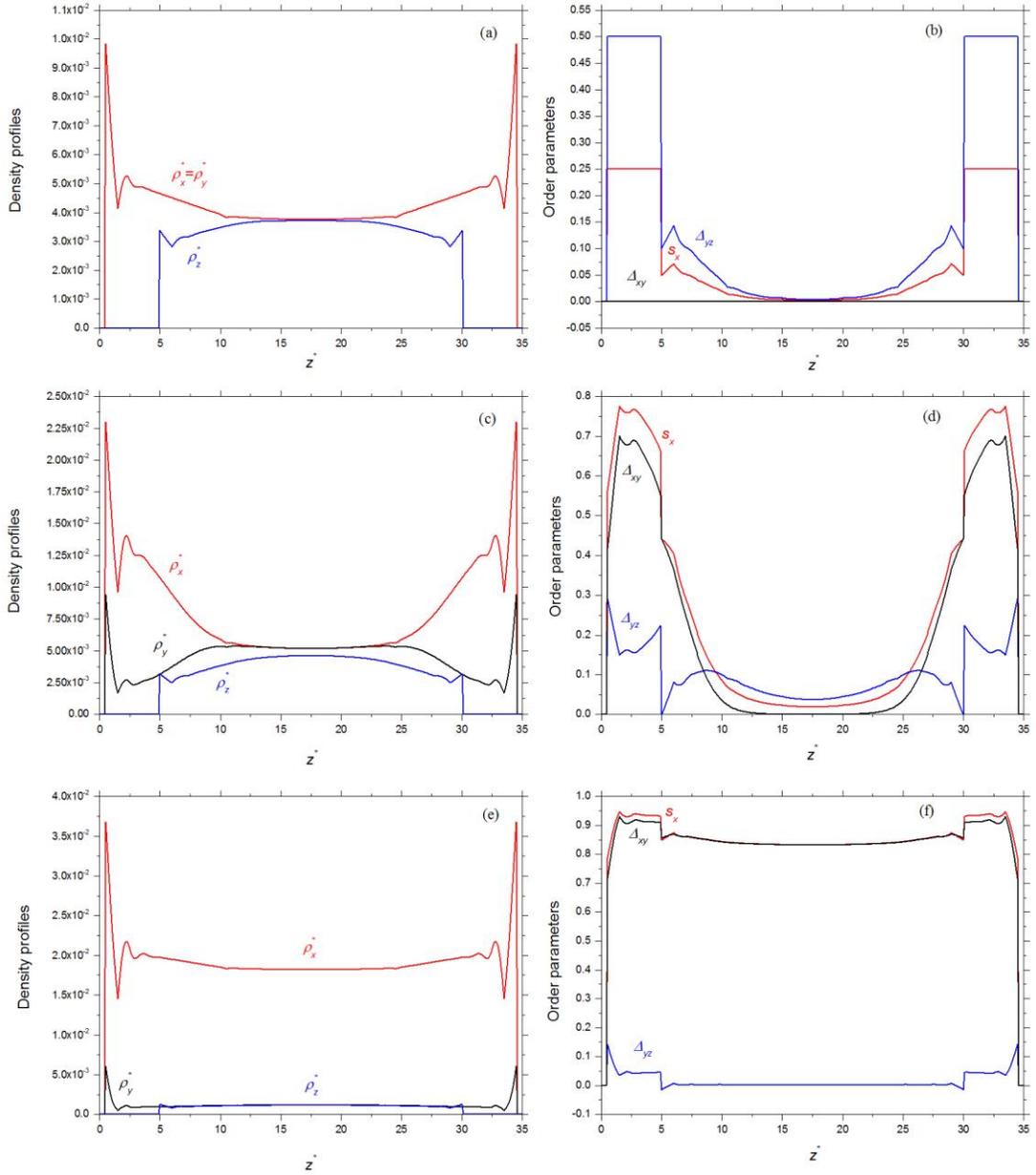

**FIG. 5.** (Color online) Density profiles and order parameters of confined hard rods ($L/D=10$) between two hard walls at reduced wall separation $h^* = H/D = 35$. Figs. (a) and (b) are obtained at $\beta\mu = -3.48$ and show that the phase is isotropic. Figs. (c) and (d) correspond to $\beta\mu = -2.51$, the phase is biaxial near the walls ($\rho_x^* \neq \rho_y^*$) but isotropic in the middle of the pore. In Figs. (e) and (f) the reduced chemical potential is -2, where the phase is biaxial nematic.

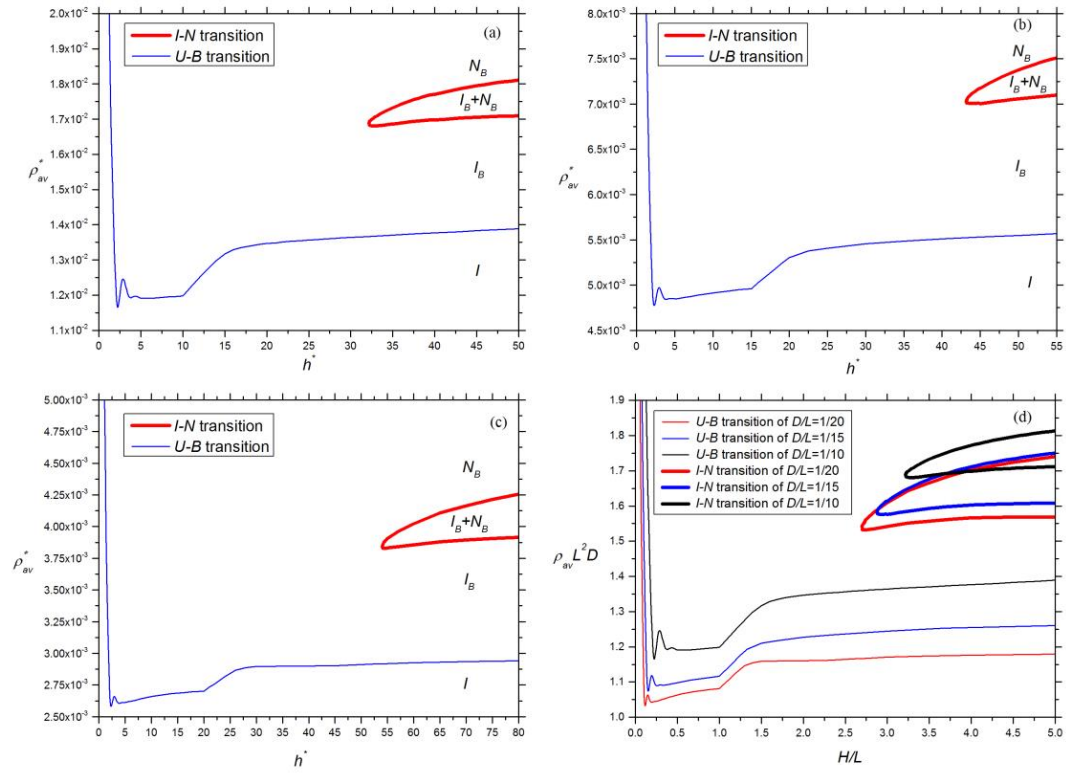

**FIG. 6.** (Color online) Phase diagram of confined hard rods in density - wall-to-wall separation plane. The $I_B$-$N_B$ (thick lines) and $U$-$B$ (thin lines) phase transition curves are shown. The values of the aspect ratios: (a) $L/D=10$, (b) $L/D=15$ and (c) $L/D=20$. The resulting phase diagrams are shown together in different units in (d).

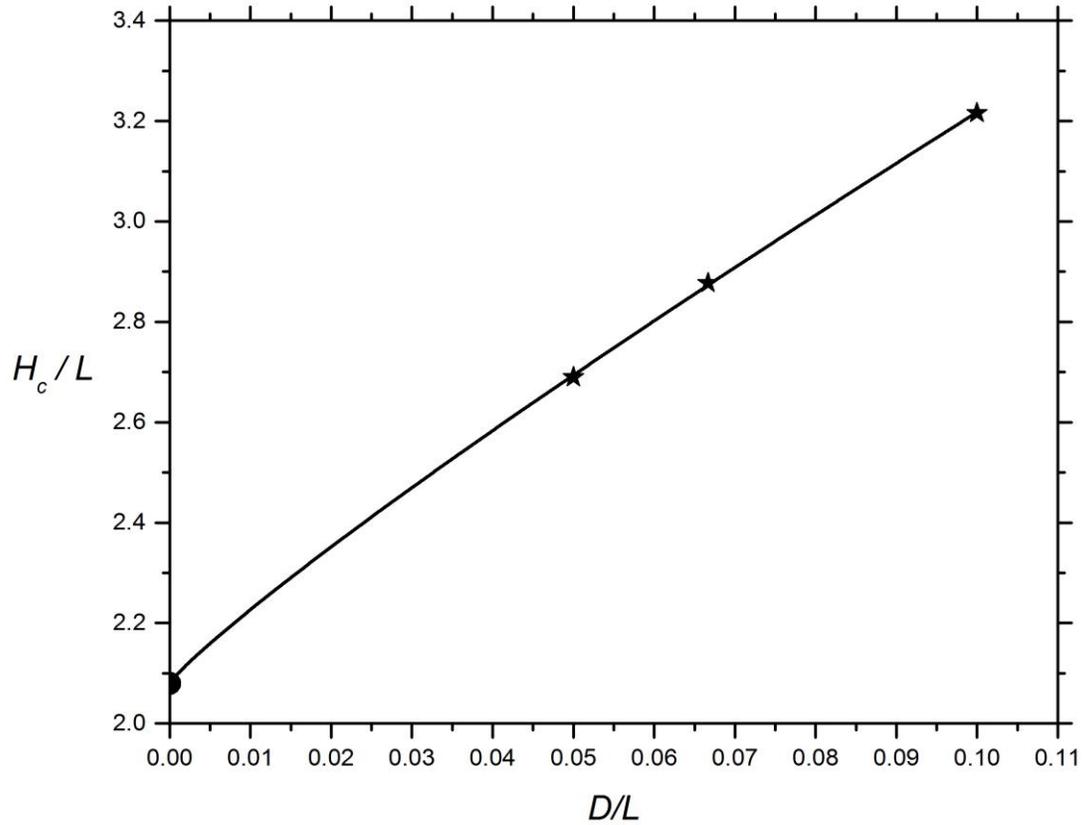

**FIG. 7**. The critical pore width ($H_c/L$) as a function of $D/L$. The star points correspond to our numerical results, while the solid circle is the result of van Roij et al. for $D/L \to 0$ [11]. The fitted equation (solid line) is given by $H_c/L = 2.08(1+ax^b)$, where $x = D/L$, $a = 4.24857$ and $b = 0.89038$.